\documentclass[useAMS,usenatbib]{mn2e}

\newcommand{\hMpc}{\, h^{-1}\, \mathrm{Mpc}}
\newcommand{\hMs}{\, h^{-1}\, \mathrm{M}_{\sun}}
\usepackage{graphicx} 
\title[Real-space correlation function from photo-z]{Recovering the real-space correlation function from photometric redshift surveys}

\author[P.~Arnalte-Mur et al.]{Pablo~Arnalte-Mur$^{1,2}$\thanks{E-mail: pablo.arnalte@uv.es}, 
Alberto~Fern\'andez-Soto$^{3}$, 
Vicent~J.~Mart\'inez$^{1,2}$, \newauthor
Enn~Saar$^{4}$,
Pekka Hein\"am\"aki$^{5}$ and 
Ivan Suhhonenko$^{4}$ \\
$^{1}$Observatori Astron\`omic,  Universitat de Val\`encia, Apartat de Correus 22085, E-46071 Val\`encia, Spain\\
$^{2}$Departament d'Astronomia i Astrof\'isica, Universitat de Val\`encia, Apartat de Correus 22085, E-46071 Val\`encia, Spain\\
$^{3}$Instituto de F\'isica de Cantabria (CSIC-UC), Avda de los Castros s/n, E-39005 Santander, Spain\\
$^{4}$Tartu Observatoorium, T\~oravere, 61602 Estonia\\
$^{5}$University of Turku, Department of Physics and Astronomy, Tuorla Observatory, V\"ais\"al\"antie 20, Piikki\"o, Finland }

\begin{document}

\date{}

\pagerange{\pageref{firstpage}--\pageref{lastpage}} \pubyear{}

\maketitle

\label{firstpage}

\begin{abstract}
Measurements of clustering in large-scale imaging surveys that make
use of photometric redshifts depend on the uncertainties in the
redshift determination. We have used light-cone simulations to show 
how the
deprojection method successfully recovers the real space correlation
function when applied to mock photometric redshift surveys. We study
how the errors in the redshift determination
affect the quality of the recovered two-point correlation function.
Considering the expected errors associated to the planned 
photometric redshift surveys, we conclude that this method 
provides information on the clustering of matter useful 
for the estimation of cosmological parameters that depend on
the large scale distribution of galaxies.
\end{abstract}

\begin{keywords}
methods: data analysis -- methods: statistical -- techniques: photometric -- galaxies: distances and redshifts -- large-scale structure of Universe
\end{keywords}

\section{Introduction}

In recent years, photometric redshift surveys have been proposed
as a way to extend large-scale structure studies towards higher 
redshifts than it is  possible using spectroscopic surveys. These 
surveys observe a region of the sky through a number of  filters, 
and use the photometry obtained to determine the redshifts, $z$, and spectral energy distributions
(SED's) of galaxies. Using photometry instead of spectra allows them to get 
much deeper, but the uncertainty in the determination of redshifts is larger \citep{bau62a, 
koo86a, con95a, bla05a, fer01a}.

Two projects of this kind are COMBO-17(Classifying Objects by Medium-Band Observations)
 and the ALHAMBRA (Advanced Large, Homogeneous Area Medium Band Redshift Astronomical) survey. 
COMBO-17 \citep{wol03a}
 surveyed a total area of $\sim 1 \deg^2$ 
using a combination of 17 broad-band and medium-band filters. It 
provided photometric redshifts for $\sim 25 000$ galaxies in $0.2 < z < 1.2$, 
with a typical error of $\Delta z \simeq 0.03$. The ALHAMBRA survey 
\citep{mol06a, mol08a, fer08a},
 currently ongoing, will observe a total area of 
$\sim 4 \deg^2$, in 16 $1\degr \times 0\fdg25$ strips. 
It uses 20 medium-band, equal-width filters covering the optical 
range, plus the standard $J$, $H$, $K_s$ near-infrared filters. \citeauthor{mol06a} 
expect to obtain photometric redshifts for $\sim 3 \times 10^5$ galaxies 
with $I_{AB} \leq 24.7$ ($60$ per cent completeness level), $z_{med} = 0.74$, 
and $\Delta z \simeq 0.015 (1+z)$.

These surveys can therefore provide us with  large and deep samples of 
galaxies. In order to study large-scale structure using these samples,
one has to deal with large redshift errors.
These redshift errors produce uncertainties in the determination of distances 
to the galaxies, and hence in their three-dimensional positions \citep{coe06a}. This 
uncertainty has to be added to the one produced by peculiar motions of 
galaxies. The latter is important for spectroscopic surveys, while the 
former dominates the uncertainties in photometric surveys.

In the present work, we focus on the two-point correlation function, $\xi(r)$. 
We study how it is affected by redshift errors, and describe a method 
to recover its real-space value from photometric redshift survey data. 
The method we use is based in measuring the two-dimensional correlation 
function, $\xi(\sigma, \pi)$ (where $\pi$ is  the line-of-sight 
separation, and $\sigma$ is the transverse separation), obtaining 
from it the projected correlation function, $\Xi(\sigma)$, and 
deprojecting it. This method (outlined in Section 3) was first proposed by \citet{dav83a} as 
a way to avoid the uncertainties due to peculiar velocities in spectroscopic 
surveys, and has been used successfully in subsequent analyses 
\citep*[e.g.][]{sau92a, haw03a, mad03a, zeh04a, zeh05a}.

\citet{phl06a} studied the correlation function of galaxies in COMBO-17. 
They used $\Xi(\sigma)$ as a measure of real-space clustering, and 
compared it to the predictions of halo occupation models. However, 
they did not attempt to recover $\xi(r)$ from their data.

We tested this method using data from the light-cone simulation of 
\citet{hei05a}. From the simulation, we produced three mock photometric 
redshift catalogues, corresponding to different accuracies in the 
determination of redshifts. We then compared the correlation function $\xi(r)$ obtained 
by our method in each case to the real-space one, computed from the original catalogue.

We describe the simulation and the way in which we created the mock 
catalogues in Section~\ref{sec:data}. Section~\ref{sec:meth} describes 
our method to calculate $\xi(r)$ from the simulated data. In Section~\ref{sec:res} 
we present our results, and we summarize our conclusions in Section~\ref{sec:conc}.

\section{Data used}
\label{sec:data}

The catalogues used in this paper come from the light-cone simulation 
of \citet{hei05a}. They simulated the distribution of dark matter haloes 
in a  light-cone covering $2\degr \times 0\fdg5$ in the sky for a 
standard $\Lambda$CDM cosmology ($\Omega_{DM} = 0.226$, $\Omega_B = 0.044$, 
$\Omega_{\Lambda} = 0.73$, 
$h = H_0 / 100 \,\mathrm{km}\, \mathrm{s}^{-1} \, \mathrm{Mpc}^{-1} = 0.71$, 
$\sigma_8 = 0.84$). We used the full output catalogue of the simulation, 
which contains haloes with $M \geq 7.14 \cdot 10^{10} \hMs$. We calculate 
the halo correlation function in our analysis. Its behaviour should be 
similar enough to the galaxy correlation function as to correctly assess 
the validity of our method.

As in this work we are not interested in the evolution of the correlation function 
with redshift, we restricted our study to the redshift bin $z \in [2,3]$. 
The volume considered, in co-moving coordinates, is $864\hMpc$ long in 
the line-of-sight direction, 
while its transverse section varies between $130 \times 32 \hMpc$ in its 
close end to $160 \times 40 \hMpc$ in its far end. The total volume is 
$4.56 \times 10^{6} \, h^{-3}\, \mathrm{Mpc}^3$ and it contains 
$\sim 180 000$ haloes. We chose that redshift interval for two reasons. 
At lower redshifts, the light-cone is too narrow, while at higher redshifts it contains 
too few haloes.

We generated three mock `photometric redshift catalogues', corresponding 
to surveys with redshift  uncertainties $\Delta z / (1+z) = 0.05, 0.015, 0.005$. 
The first case ($\Delta z /(1+z) = 0.05$) corresponds typically to a 
survey using $\sim5$ broad-band filters \citep[see e.g.][]{fer01a}. 
$\Delta z / (1+z) = 0.015$ corresponds to the value expected from the ALHAMBRA 
survey \citep{mol08a}. The last case, $\Delta z / (1+z) = 0.005$, would correspond to a 
future survey using even more filters. 
As an example, the PAU (Physics of the Accelerating Universe) survey project \citep{ben08a} 
aims at obtaining photometric redshifts for Luminous Red Galaxies (LRG's) with uncertainties $\Delta z/(1+z) \sim 0.0035$ for 
$z \la 0.9$. As the uncertainty in photometric redshifts decreases for high redshift galaxies ($z \ga 2.5$), when the Lyman-$\alpha$ 
wavelength enters into the visible domain, it should also be possible, in principle, to get such a small $\Delta z$ in this case.

In creating our mock catalogues, we assumed Gaussian errors for the photometric redshifts. This is not
generally the case for real surveys, due to the existence of a fraction of `catastrophic' redshift determinations, 
and to the mix of different classes of objects with a variety of photometric redshift errors. However, our assumption
of single-peaked Gaussian-distributed errors would be valid for a catalogue selected to contain only ``good'' redshifts. 
This catalogue could be built combining the selection of a given class of objects (e.g. LRG's), with the use of some estimate
of the redshift determination quality. 
The latter could be the knowledge of the full redshift probability function \citep{fer02a}, 
or the `odds' parameter in the case of Bayesian methods \citep{ben00a}. Existing experience indicates that, 
depending on the survey design, it is possible to obtain ``good'' redshifts for objects down to magnitudes
$m_{\rm lim} - 1$ or $m_{\rm lim} - 2$, where $m_{\rm lim}$ is the limit magnitude of the survey.

At the end of Section~\ref{ssec:tests}, we assess the robustness of our results to the presence of `catastrophic'
redshifts. We consider catalogues with a fraction of such outliers of 5 per cent. This is a conservative value, typical of broad-band, 
non-optimized photometric redshift surveys, and it should be significatively smaller in the case of ``good'' redshifts. 
As an example, \citet{ilb08a} compiled a photometric redshift catalogue for the COSMOS field, using 30 bands ranging from the ultraviolet to the 
mid-infrared. They obtained just $0.7$ per cent of outliers when comparing their bright sample ($i^+_{AB} < 22.5$) to spectroscopic 
redshifts.

To generate each mock catalogue, we modified the position of each point 
in the simulation following these steps:
\begin{enumerate}
\item We calculated the `cosmological redshift' of the object from its 
real-space position.
\item We added to this `cosmological redshift' the redshift due to the 
peculiar velocity of the object.
These peculiar velocities of the haloes are provided by the simulation.
\item To simulate the expected redshift errors, we added a random shift 
to the resulting redshift, following a Gaussian distribution with variance 
equal to $\Delta z$ in each case. The redshift obtained is the `observed redshift' of the object.
\item We finally obtained the three-dimensional position of the object 
corresponding to this `observed redshift' and included it in the mock catalogue.
\end{enumerate}

This distortion process was carried out for all the points in the whole cone of
the simulation. The selection of the points in the redshift bin $z \in [2,3]$ was performed
using the new `observed redshifts', thus simulating the selection process in  a
real survey.

\begin{figure*}
\begin{minipage}{168mm}
\includegraphics[width=\textwidth]{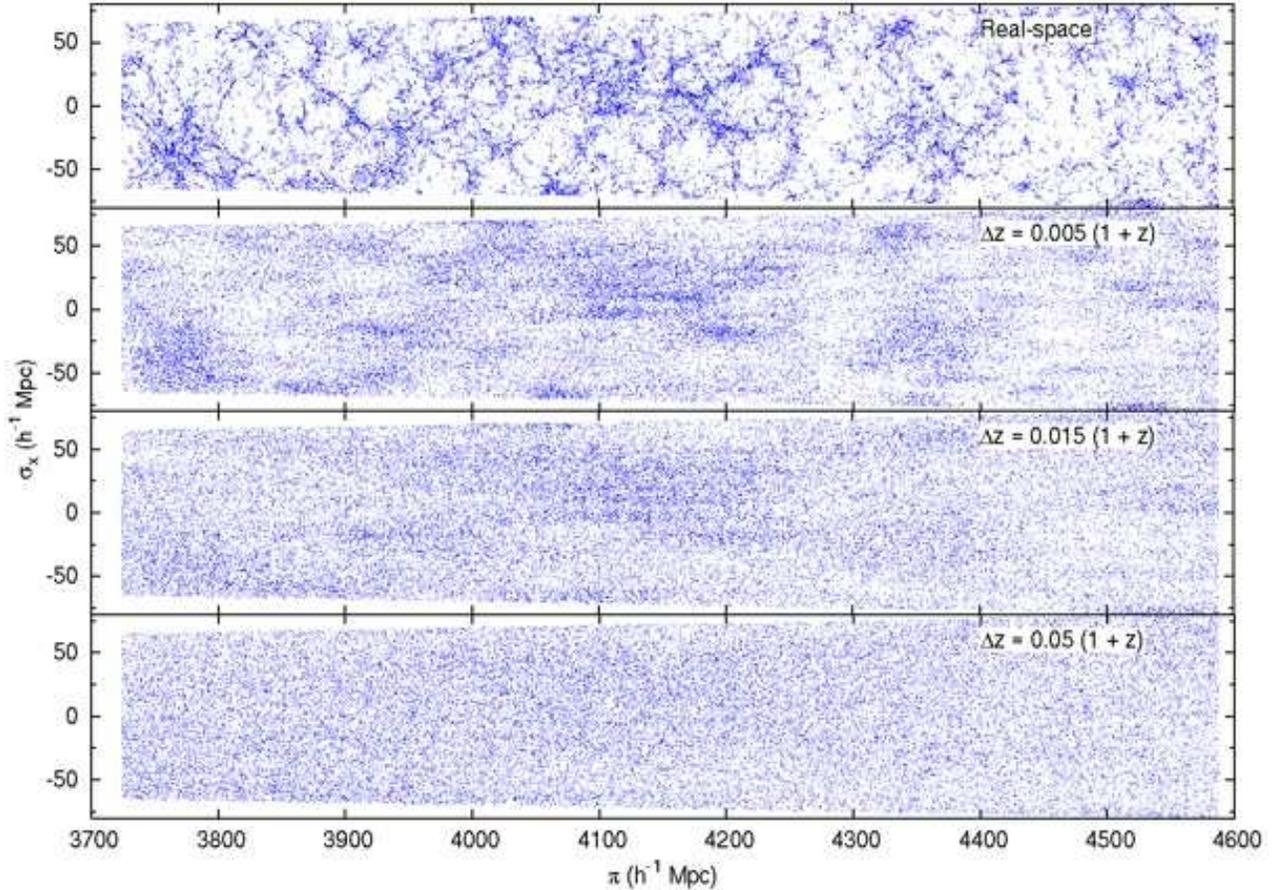}
\caption{The distribution of haloes in the four catalogues used: the original 
real-space catalogue, and the three mock photometric catalogues. The distribution 
is shown projected on a longitudinal plane, and only $20$ per cent of the points 
are shown, for clarity.}
\label{fig:cats}
\end{minipage}
\end{figure*}

Fig.~\ref{fig:cats} shows the distribution of haloes in the original 
catalogue and in the mock photometric catalogues. 
The Figure shows the real space positions of haloes, not affected
by peculiar velocities, and thus 
does not show the finger-of-God or coherent infall effects observed in spectroscopic 
surveys.
Due to redshift errors, 
structures which are clearly seen in real space are smoothed and hardly 
recognizable in photometric redshift data.

%%%%%%%%%%%%%%%%
%%%%%%%%%%%%%%%%%

\section{Method}
\label{sec:meth}

The most widely used method to measure the correlation function consists in 
comparing the distribution of points in the data catalogue with a 
random distribution of points generated in the same volume. To make 
the comparison, one calculates the number of pairs with separation 
in the range $[r, r + \mathrm{d}r]$ between points in the data catalogue ($DD(r)$), 
between points in the random catalogue ($RR(r)$), and between a point in the real 
catalogue and a point in the random catalogue ($DR(r)$).  The estimator used in this work 
to compute $\xi(r)$ is \citep{lan93a}:
\begin{equation}
\label{eq:LS}
\hat{\xi}(r) = 1 + \left( \frac{N_R}{N_D} \right) ^2 \frac{DD(r)}{RR(r)} - 
2 \frac{N_R}{N_D} \frac{DR(r)}{RR(r)} \, ,
\end{equation}
where $N_D$ is the number of points in the data catalogue, and $N_R$ is the
number of points in the random catalogue.

However, this method can not be used when the studied catalogue comes from 
a photometric survey. The large errors in redshift and hence in the 
line-of-sight positions produce two effects that have to be considered. 
On one side, these random 
shifts in position erase correlations between points, and hence 
$\xi(r)$ measured according to (\ref{eq:LS}) would be much lower than 
the real $\xi(r)$. On the other side, as the shifts are only in the 
line-of-sight direction, isotropy of the distribution is lost. 
Correlation is only lost along the longitudinal direction, but it is 
conserved in the transverse plane.

The method we used to recover real-space $\xi(r)$ from the mock 
photometric redshift catalogues is the same described in \citet{dav83a} 
and \citet{sau92a} for spectroscopic surveys. It is based in the 
decomposition of pair separations in parallel and perpendicular distances 
($\pi$ and $\sigma$, respectively). 

Let $\bmath{s_1}$ and $\bmath{s_2}$ be the measured positions (in `observed 
redshift space') of two points in the catalogue. We then define the 
separation vector, $\bmath{s} \equiv \bmath{s_2} - \bmath{s_1}$, and 
the line-of-sight vector, $\bmath{l} \equiv \bmath{s_1} + \bmath{s_2}$, 
of the pair. From these, we now define the parallel and perpendicular distances of the pair as
\begin{equation}
\label{eq:parperp}
\pi \equiv \frac{|\bmath{s} \cdot \bmath{l}|}{|\bmath{l}|} \quad , 
\quad \sigma \equiv \sqrt{\bmath{s} \cdot \bmath{s} - \pi ^2 } \, .
\end{equation}

Once we have defined $\pi$ and $\sigma$ for each pair of points, we can 
calculate the two-dimensional correlation function, $\xi(\sigma, \pi)$ 
in an analogous way to equation~(\ref{eq:LS}), substituting the $(r)$ 
dependence by $(\sigma, \pi)$. From $\xi(\sigma, \pi)$, we define the 
projected correlation function as
\begin{equation}
\label{eq:wp}
\Xi (\sigma) \equiv 2 \int_0^{\infty} \xi(\sigma, \pi)  \mathrm{d}\pi \, .
\end{equation}
As $\Xi$ depends only on $\sigma$, and the angle between any pair of points 
is small, it will not be affected significantly by redshift errors, 
as these will mainly produce shifts in $\pi$.

Assuming that the real-space distribution is isotropic, we can relate 
$\Xi$ to the real-space correlation function, $\xi_r$, as
\begin{equation}
\label{eq:isotropy}
\Xi (\sigma) = 2 \int_{\sigma}^{\infty} \xi_r(r) 
\frac{r \mathrm{d}r}{\left( r^2 - \sigma^2 \right) ^{1/2}} \, .
\end{equation}
This relation can be inverted, 
obtaining $\xi_r$ in terms of $\Xi$ as the Abel integral:
\begin{equation}
\label{eq:wp2xi}
\xi_r (r) = - \frac{1}{\upi} \int_{r}^{\infty} 
\frac{\mathrm{d}\Xi(\sigma)}{\mathrm{d}\sigma} 
\frac{\mathrm{d}\sigma}{\left( \sigma^2 - r^2 \right) ^{1/2}} \, .
\end{equation}

Therefore, the method proposed to compute $\xi(r)$ from  photometric 
survey data consists of the following steps. We first obtain $\xi(\sigma, \pi)$ from counting 
pairs of points in the data and in the random catalogues. The projected 
correlation function, $\Xi (\sigma)$, is then obtained by 
integration of equation~(\ref{eq:wp}). Finally, the real-space 
correlation function, $\xi(r)$, is calculated from equation~(\ref{eq:wp2xi}). 
Some problems arise in the numerical integration of equations 
(\ref{eq:wp}) and (\ref{eq:wp2xi}). Both integrals extend formally to $+ \infty$. However,
when computing them numerically, we have to set finite upper limits, $\pi_{\rm max}$ and 
$\sigma_{\rm max}$.

In the first case, the value of $\pi_{\rm max}$ should be 
large enough to include almost all the correlated pairs. However, if 
it is too large, this would introduce extra noise in the calculation. 

When integrating equation~(\ref{eq:wp2xi}), the upper limit 
$\sigma_{\rm max}$ is fixed, for pencil-beam surveys, by the maximum 
transverse separation allowed by the geometry. The way we used to 
evaluate  (\ref{eq:wp2xi}) was that of \citet{sau92a}. We interpolated 
linearly $\Xi$ between its values in each $\sigma$ bin, and then integrated (\ref{eq:wp2xi})
analytically. Taking $\Xi_i$ 
as the value of $\Xi$ for the bin centred at $\sigma_i$, we have
\[
\xi( \sigma_i) = - \frac{1}{\upi} \sum_{j \geq i} 
\frac{\Xi_{j+1} - \Xi_{j}}{\sigma_{j+1} - \sigma_{j}} 
\ln \left( \frac{\sigma_{j+1} + \sqrt{\sigma_{j+1}^2 - \sigma_i^2}}
{\sigma_{j} + \sqrt{\sigma_j^2 - \sigma_i^2}} \right)  \, .
\]

Redshift errors will influence the result in two ways. First, these errors
change the apparent line-of-sight direction $\bmath{l}/|\bmath{l}|$
(see equation (\ref{eq:parperp})), and through that, the apparent line-of-sight
distance $\pi$, and, most important, the perpendicular distance $\sigma$.
These errors grow with the redshift error and with the width of the galaxy
pair.

Another, and much stronger source of errors is the assumption that the
apparent  distance in redshift space is the real distance
 between two
galaxies -- this assumption is necessary to obtain our basic integral
relation (\ref{eq:isotropy}). In case of photometric errors, this assumption
is hardly justified, but we will see that the inverted correlation functions are
close to the real one, anyway. The errors caused by this assumption grow with 
the redshift errors.

\section{Results}
\label{sec:res}

We tested the deprojection method to measure $\xi(r)$ using the 
mock photometric catalogues described in Section~\ref{sec:data}. 
We applied the method described above to data from the three mock 
catalogues, and obtained $\xi_{\rm dep}(r)$ in each case. These 
were then compared to the real-space $\xi_{r}(r)$ calculated  
from the undistorted simulation catalogue using equation~(\ref{eq:LS}) directly. 

The value used  for the integration limit in equation~(\ref{eq:wp2xi}) 
was $\sigma_{\rm max} = 130 \hMpc$. This is about $80$ per cent of 
the maximum transverse separation allowed by the geometry of the light-cone. 
We used $32$ bins in $\sigma$, logarithmically spaced between $0.1 \hMpc$ 
and $\sigma_{\rm max}$.

We performed several tests to choose the appropriate value for $\pi_{\rm max}$, scaling it relative to the
$\Delta z$ of the catalogue considered. The $\xi(r)$ recovered increased with $\pi_{max}$ for values
$\pi_{\rm max} \la 3 \Delta z$, and converged to fixed values for larger $\pi_{\rm max}$. However, the noise
also increased with $\pi_{\rm max}$. We adopted a conservative value of $\pi_{\rm max} \simeq 4\Delta z$ for our
calculations, in order to be sure to include all correlated pairs, and not introducing too much extra noise.

To estimate the correlation function error and covariance between
bins in $r$, we used the jackknife method \citep[see e.g.][]{zeh05b}.
We divided our volume in $12$ equal sub-volumes, and constructed our
jackknife samples omitting one sub-volume at a time. We repeated the full calculation
of $\xi(r)$ for each of these samples. Denoting by $\xi_i^k$ the 
value of the correlation function obtained for bin $i$ in jackknife sample $k$, 
the covariance matrix is then
\[
C_{ij} = \frac{N - 1}{N} \sum_{k=1}^{N} \left( \xi_i^k - \bar{\xi_i} \right)  \left( \xi_j^k - \bar{\xi_j} \right) \, , 
\]
where $\bar{\xi_i}$ is the average of the values obtained for bin $i$, and $N=12$.

\subsection{Effect of redshift errors on \mbox{\boldmath{$\xi(\sigma,\pi)$}}}
\label{ssec:xi2d}

\begin{figure*}
\begin{center}
\includegraphics[scale=1]{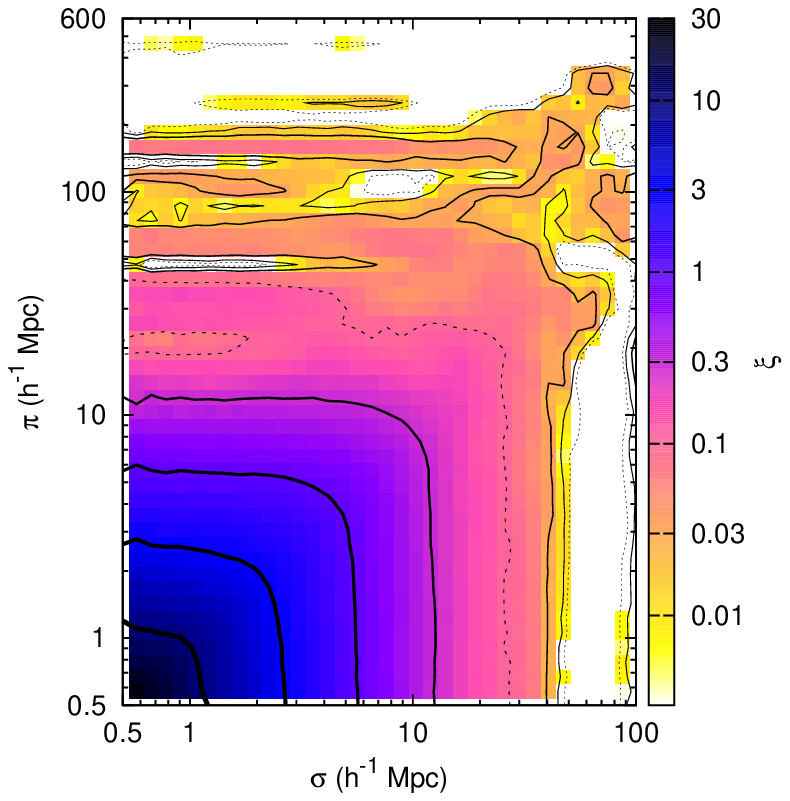}
\includegraphics[scale=1]{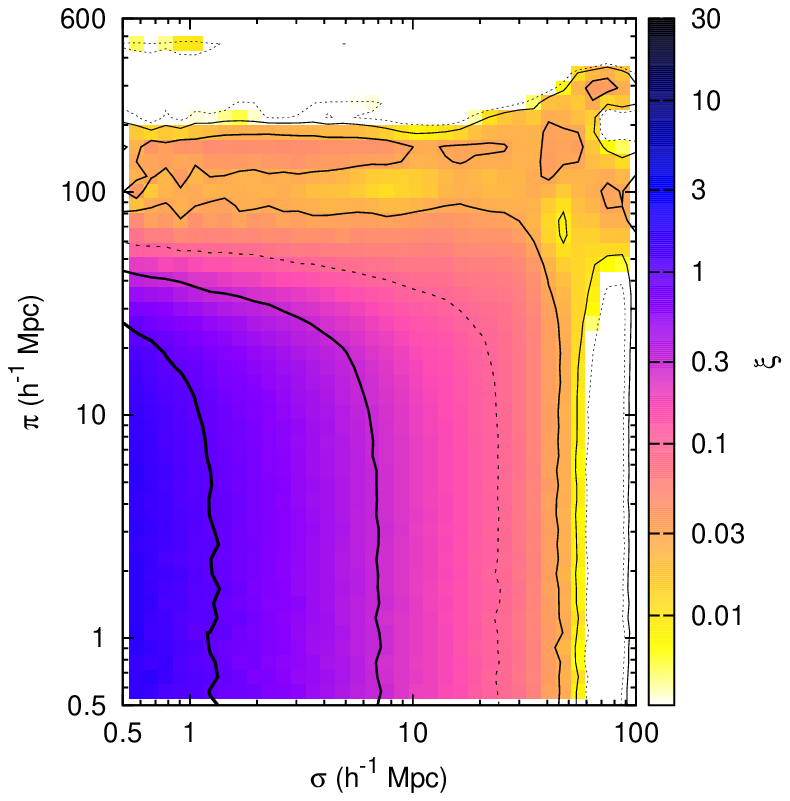}
\includegraphics[scale=1]{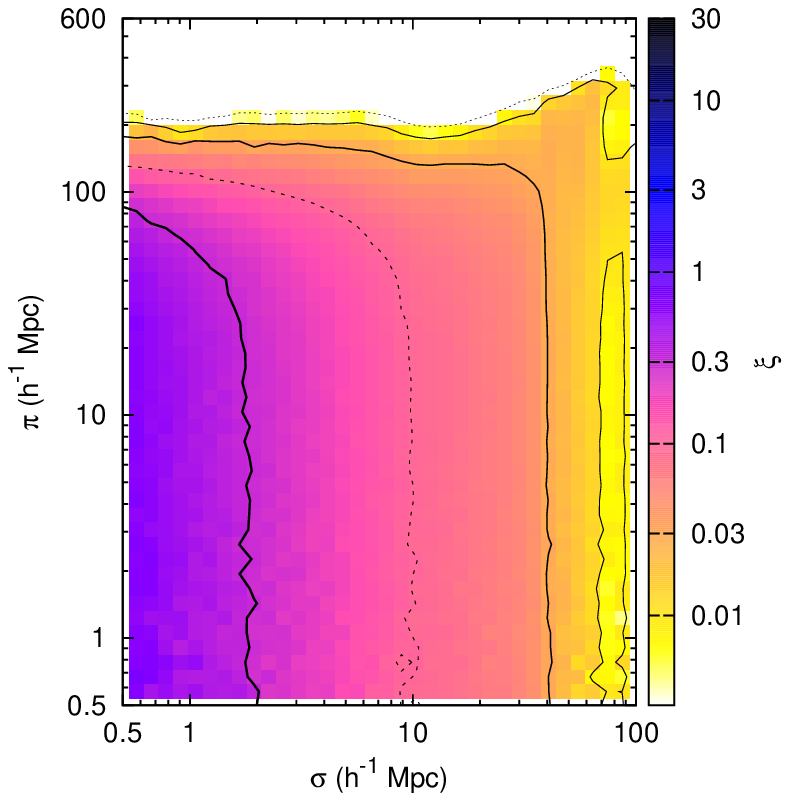}
\includegraphics[scale=1]{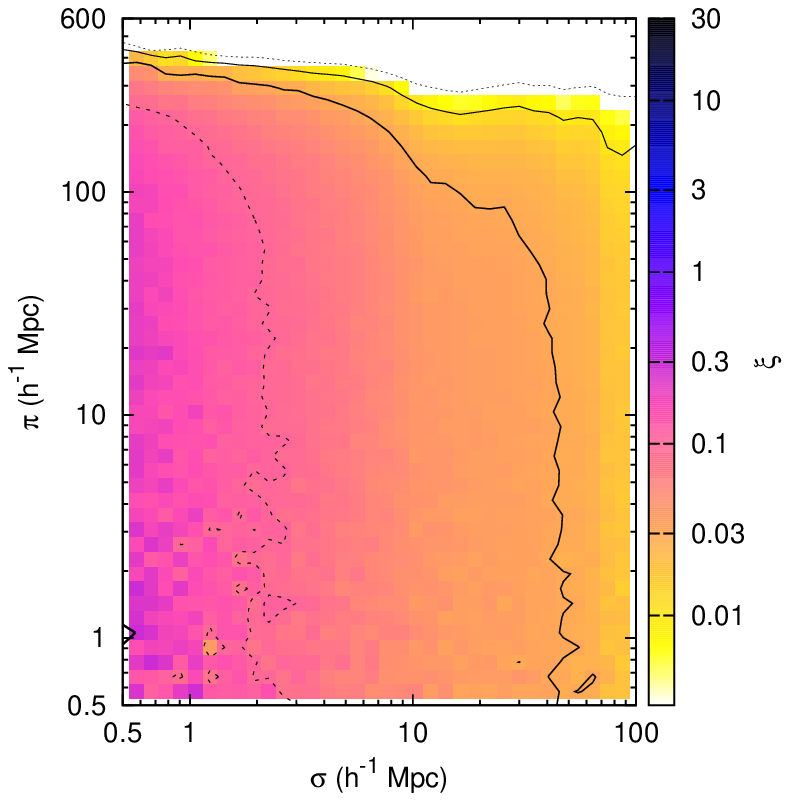}
\end{center}
\caption{The two-dimensional correlation function $\xi(\sigma,\pi)$ obtained for the real space
catalogue (top left), and for 
the mock photometric catalogues with $\Delta z = 0.005(1+z)$ (top right), 
$\Delta z = 0.015(1+z)$ (bottom left) and $\Delta z = 0.05(1+z)$ (bottom right). Contours are 
drawn at $\xi = 10, 3, 1, 0.3, 0.1, 0.03, 0.01, 0$, with decreasing thickness. 
Contours at $0.1$ and $0$ are dashed.}
\label{fig:plot2d}
\end{figure*}

As a first step in the calculation of $\xi(r)$, we calculated $\xi(\sigma,\pi)$ 
for each mock catalogue. The results are shown in Fig.~\ref{fig:plot2d}. We also
plot the $\xi(\sigma, \pi)$ obtained in the real-space catalogue, for comparison.
Two effects of the redshift errors can be observed. First, correlation decreases
with the value of $\Delta z$ for each catalogue. Also, there is a loss of symmetry of 
$\xi(\sigma,\pi)$ in these plots. In real space, due to the isotropy of the distribution,
$\xi(\sigma,\pi)$ has circular symmetry (seen as a `boxy' symmetry in the logarithmic scale used). 
However, we can see that when we calculate it for the mock photometric catalogues, the distribution 
gets stretched along the $\pi$ axis.

%%%%%%%%%%%%%%%%
%%%%%%%%%%%%%%%%%

\subsection{Tests of the deprojection method}
\label{ssec:tests}

When comparing our results for $\xi_{\rm dep}(r)$ with the real-space 
$\xi_{r}(r)$, we restricted the analysis to the range $r \in [0.5, 30] \hMpc$. 
The lower limit is given by the way haloes were selected in the simulation. 
They were selected using a friends-of-friends algorithm, therefore if we had 
two haloes at a too small separation, they would merge into a single halo 
\citep{hei05a}. This prevents us from measuring $\xi(r)$ at such small 
distances. The upper limit was fixed because of the geometry of the light-cone. 
As the maximum separation along the short transverse axis is between $30-40 \hMpc$, 
we can not trust our method beyond these scales.

We measured $\xi_{r}(r)$ directly from the real-space catalogue.
In order to compare the real-space result to the one obtained using our method in the 
mock photometric catalogues, we fitted $\xi_{r}(r)$ by a third-order spline.

\begin{figure}
\includegraphics[scale = 1]{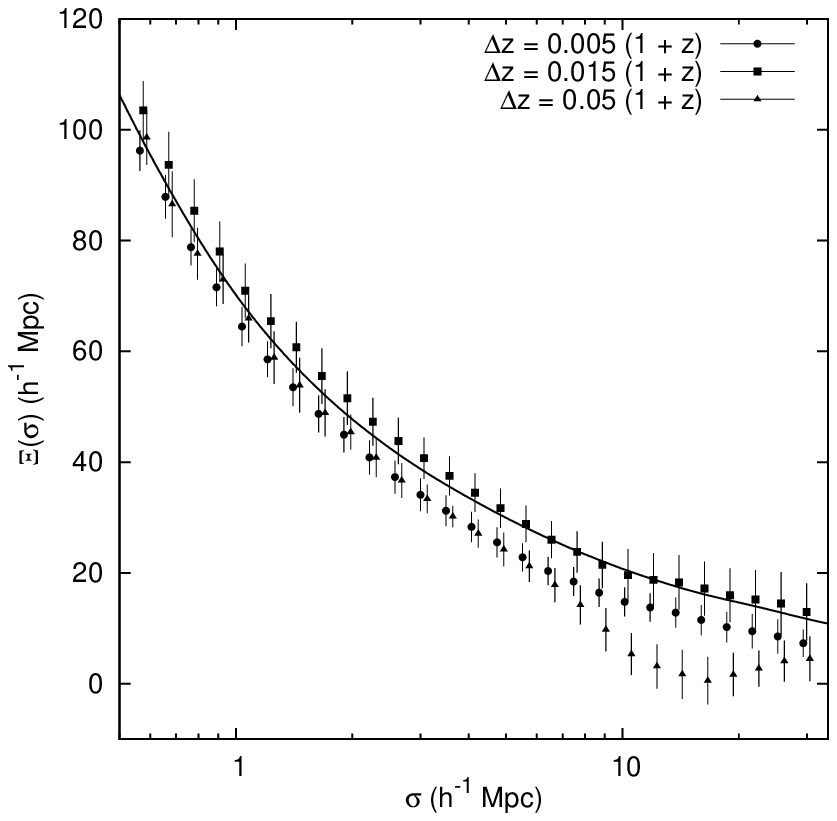}
\caption{The projected correlation function obtained from the mock photometric 
catalogues, compared to the real-space result. The solid line corresponds to a 
spline fit to $\xi_r(r)$, as explained in the text. Small shifts have been 
applied along the $\sigma$ axis, for clarity.
The feature observed at large scales for the $\Delta z = 0.05(1+z)$ catalogue
is due to the large $\pi_{\rm max}$ value used in that case.
}
\label{fig:wp}
\end{figure}

From $\xi(\sigma,\pi)$, we obtained the projected correlation function $\Xi (\sigma)$ 
for each of the mock photometric catalogues. In Fig.~\ref{fig:wp} we compare 
the function $\Xi (\sigma)$ calculated for the mock catalogues  to the function obtained 
from the spline fit to the real-space $\xi_{r} (r)$, according to equation~(\ref{eq:isotropy}).
The results 
obtained for the $\Delta z / (1+z) = 0.005$ and $\Delta z / (1+z) = 0.015$ 
catalogues follow closely the real-space result. In the case of the 
$\Delta z / (1+z) = 0.05$ catalogue, however, $\Xi (\sigma)$ falls below the 
real-space result for $r \ga 10 \hMpc$.
This feature is due to the fact that the value of $\pi_{\rm max} = 600 \hMpc$ used for that catalogue 
gets close to the line-of-sight length of the simulation box used. As $\pi_{\rm max}$
scales with $\Delta z$, this issue does not affect the other catalogues.

%%%%%%%%%%%%%%%%%%%%%
%%%%%%%%%%%%%%%%%%%%%

\begin{figure*}
\begin{center}
\includegraphics[scale = 1]{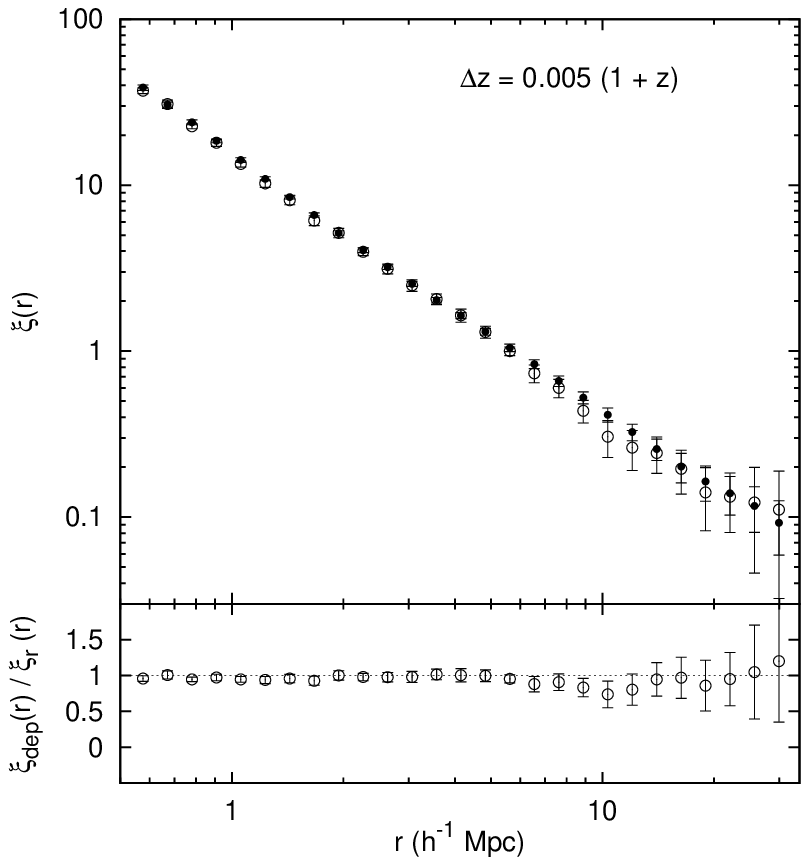}
\includegraphics[scale = 1]{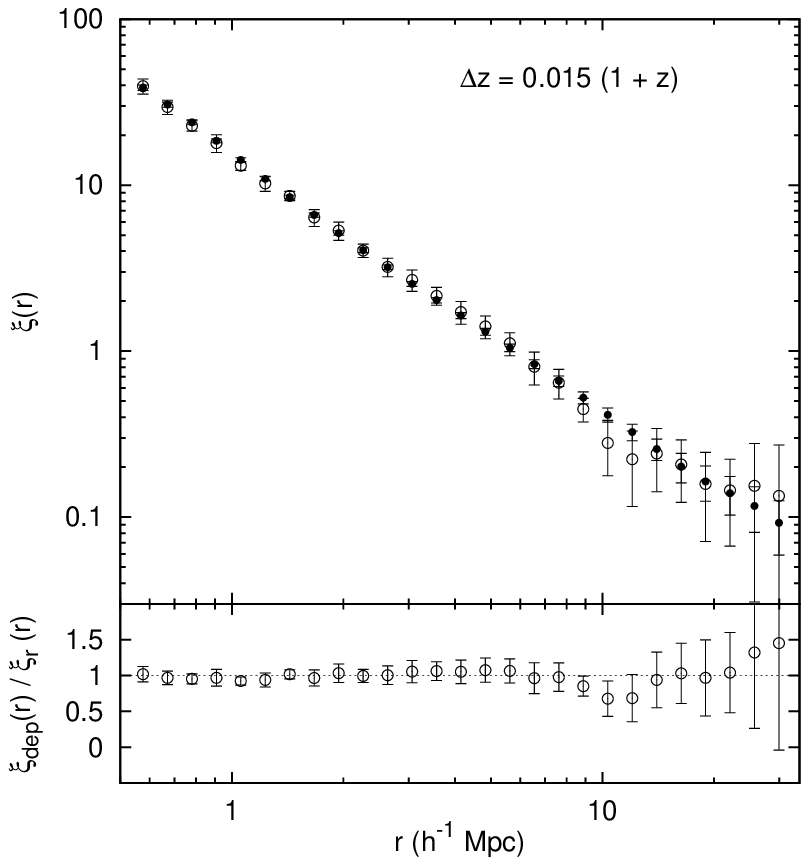}
\includegraphics[scale = 1]{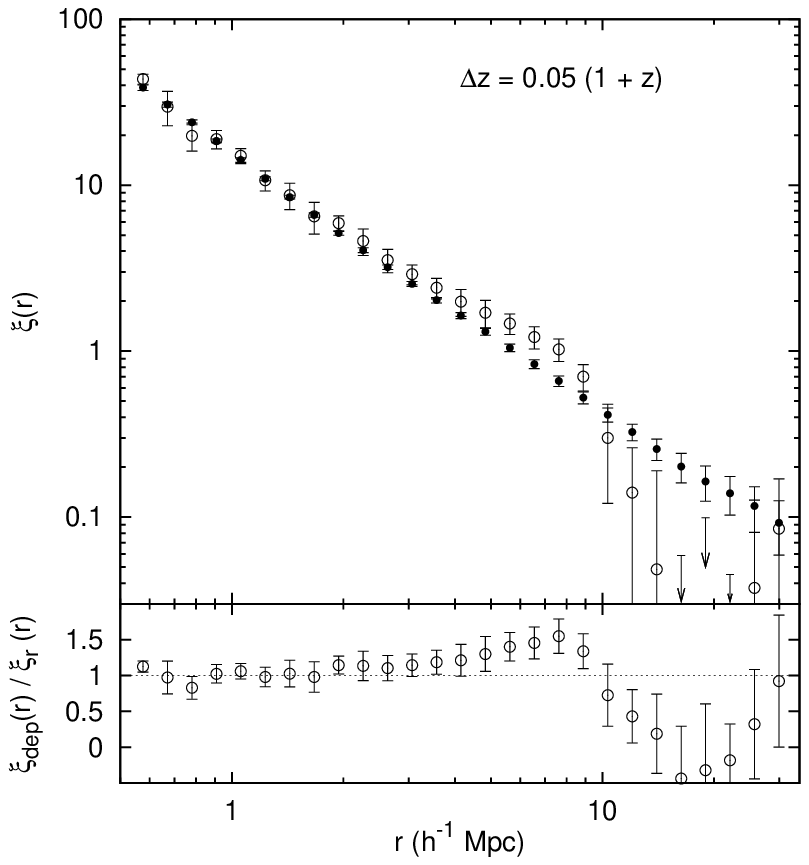}
\end{center}
\caption{Comparison between the deprojected correlation function, 
$\xi_{\rm dep}(r)$ (open circles),  and the real-space correlation 
function, $\xi_{\rm r}(r)$ (solid circles), for each mock photometric 
catalogue. The error bars
plotted correspond to the diagonal terms in the covariance matrix, $C_{ii}^{1/2}$.}
\label{fig:xidep}
\end{figure*}

Our final result for the deprojected correlation function, $\xi_{\rm dep}(r)$ 
obtained from the mock photometric catalogues is shown in Fig.~\ref{fig:xidep}, 
where we compare it to the real-space correlation function, $\xi_r(r)$. 

%%%%%%%%%%%%%%%%
%%%%%%%%%%%%%%%%%

To quantify the quality of the recovery, we used an `average normalized residual', 
$\Delta \xi$, as figure of merit, defined as
\[
\Delta \xi = \frac{1}{N} \sum_i \left\vert \frac{\xi_{\rm dep}(r_i) - \xi_{r}(r_i)}{\xi_{r}(r_i)} \right\vert \, ,
\]
where $r_i$ are the values of the bins in $r$ where we measure $\xi$, and $N$ is the number 
of such bins considered. 

Without prior knowledge of $\xi_r(r)$ we could anyhow estimate the quality of the recovery
calculating the quantity:
\[
\widehat{\Delta \xi} = \frac{1}{N} \sum_i \frac{C_{ii}^{1/2}}{\left\vert \xi_{\rm dep}(r_i) \right\vert} \, .
\]

We show the values of $\Delta \xi$ and $\widehat{\Delta \xi}$ obtained for the different mock catalogues in
Table~\ref{tab:Dxi}. We computed them for different ranges in $r$, in order to assess the validity of the 
method at different scales. From the values of $\Delta \xi$ we see that we recover $\xi(r)$ within a $5$ per cent
in the average for scales $r < 10 \hMpc$ for mock catalogues with $\Delta z \leq 0.015 (1 + z)$. 
At larger scales, the deviations from $\xi_r$ are larger ($12- 20$ per cent). In the case with the largest
redshift errors, our method is only valid for very small scales, $ r < 2 \hMpc$, where the deviations are of a 
$7$ per cent. 
We note that, in all cases where the method is valid, $\widehat{\Delta \xi} > \Delta \xi$. 
Hence, the jackknife method allows us to estimate the errors to an
acceptable precision. We remark, however, that for large values
of $\Delta z$ the jackknife error underestimates the real one as
measured from the residuals or compared to other $\Delta z$ values,
specially over medium scales ($2-20 \hMpc$).

%%%%%%%%%%%%%%%%%%%%%%%%%%5
\begin{table*}
\caption{Values of $\Delta \xi$ and $\widehat{\Delta \xi}$ obtained for the three 
			mock photometric 	catalogues and for different scale ranges.}
\label{tab:Dxi}

\begin{tabular}{@{}lcccccc} \hline
                                      & \multicolumn{2}{c}{$\frac{\Delta z}{(1 + z)}= 0.005 $} & 
										 \multicolumn{2}{c}{$\frac{\Delta z}{(1 + z)}= 0.015 $} & 
                                             \multicolumn{2}{c}{$\frac{\Delta z}{(1 + z)}= 0.05 $} \\ 
Range ($h^{-1}\, \mathrm{Mpc}$) & $\Delta \xi$ & $\widehat{\Delta \xi}$ & $\Delta \xi$ & $\widehat{\Delta \xi}$ & $\Delta \xi$ & $\widehat{\Delta \xi}$ \\ \hline
$0.5 < r  < 30$ &        $0.07$  & $0.17$       &         $0.09$    & $0.26$  &        $0.36$   &  $0.67$      \\
$0.5 < r < 2$   &         $0.04$  & $0.05$       &         $0.04$    & $0.10$  &         $0.07$   & $0.15$      \\
$2 < r < 10$ &            $0.05$  & $0.09$        &        $0.05$    & $0.16$   &        $0.28$     & $0.16$     \\
$10 < r < 30 $    &     $0.12$   & $0.40$       &         $0.20$     & $0.57$  &        $0.79$   &   $1.89$     \\ \hline
\end{tabular}
\end{table*}
%%%%%%%%%%%%%%%%%%%%%%%%%%%%%

\begin{figure}
\includegraphics[scale = 1]{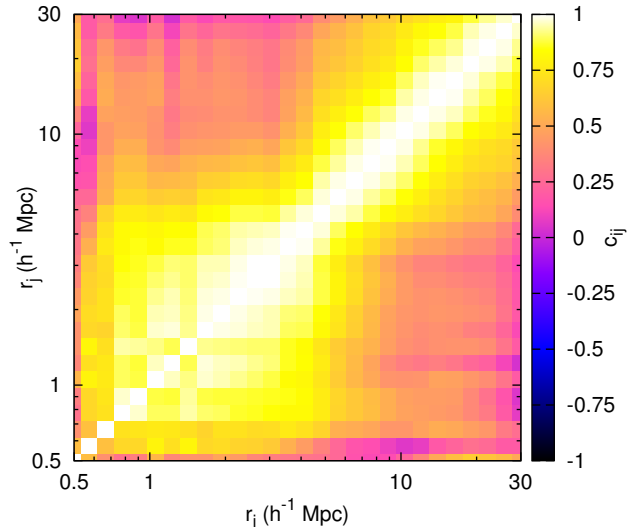}
\caption{The normalized covariance matrix ($c_{ij}$) 
of the correlation function measured directly from the real-space catalogue.}
\label{fig:real_covmat}
\end{figure}

\begin{figure*}
\begin{center}
\includegraphics[scale = 1]{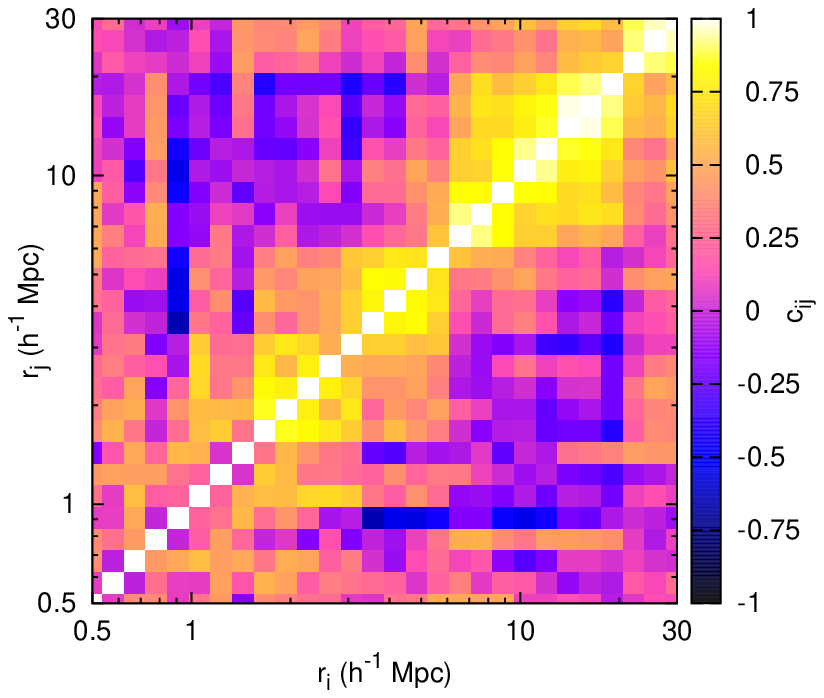}
\includegraphics[scale = 1]{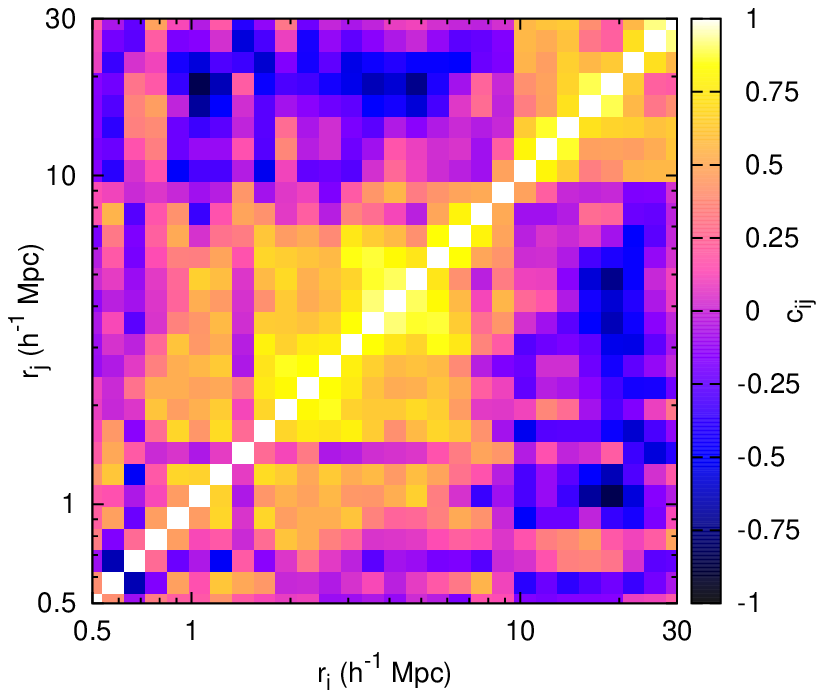}
\includegraphics[scale = 1]{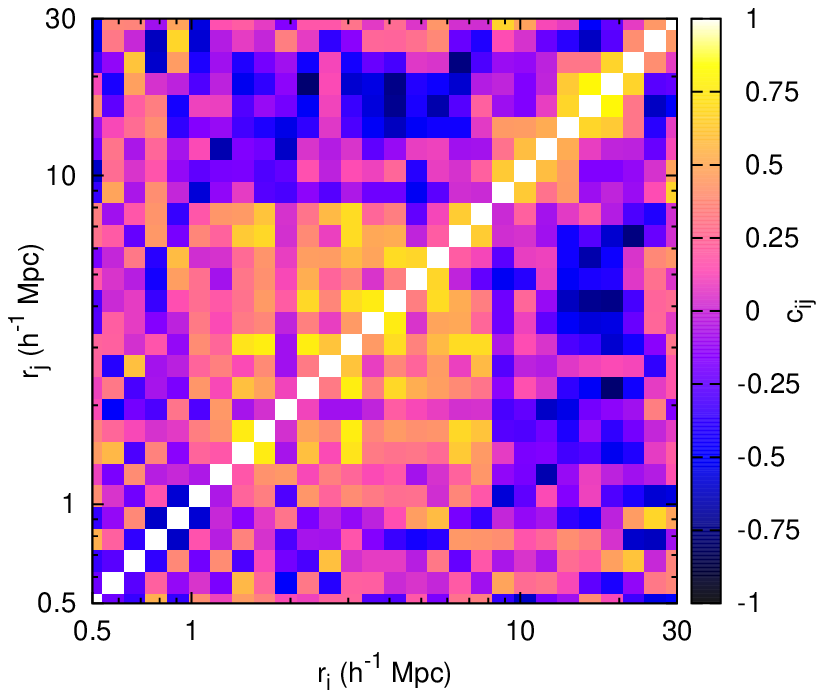}
\end{center}
\caption{The normalized covariance matrices ($c_{ij}$)  
of the deprojected 
correlation function measured directly from the mock photometric catalogues 
with $\Delta z = 0.005(1+z)$ (top left), 
$\Delta z = 0.015(1+z)$ (top right) and $\Delta z = 0.05(1+z)$ (bottom).}
\label{fig:covmat}
\end{figure*}

Fig.~\ref{fig:real_covmat} shows the covariance matrix for the real-space calculation of
$\xi_{r}(r)$, and Fig.~\ref{fig:covmat} shows it for the calculation of $\xi_{\rm dep}(r)$ in each case.
As the absolute values of the covariance drop rapidly with distance, we
show here the normalized covariances
\[
c_{ij}=\frac{C_{ij}}{\sqrt{C_{ii}C_{jj}}}.
\]

While the absolute values of the covariances grow with the redshift error, and
are always larger than the covariances for the real-space correlation function,
the structure of the covariance matrix is different. While the real-space
covariances are large for almost all bin pairs, the covariance matrices 
for photometric correlation functions are much closer to diagonal. This is
similar to the fact that the best estimates of the correlation function
are obtained integrating over line-of-sight distances, even from 
spectroscopic redshift catalogues (see, e.g., \citet{zeh05a}).

%%%%%%%%%%%%%%%%%%%%%
%%%%%%%%%%%%%%%%%%%%%

%%%%%%%%%%%%%%%%
%%%%%%%%%%%%%%%%%

\begin{figure}
\includegraphics[scale = 1]{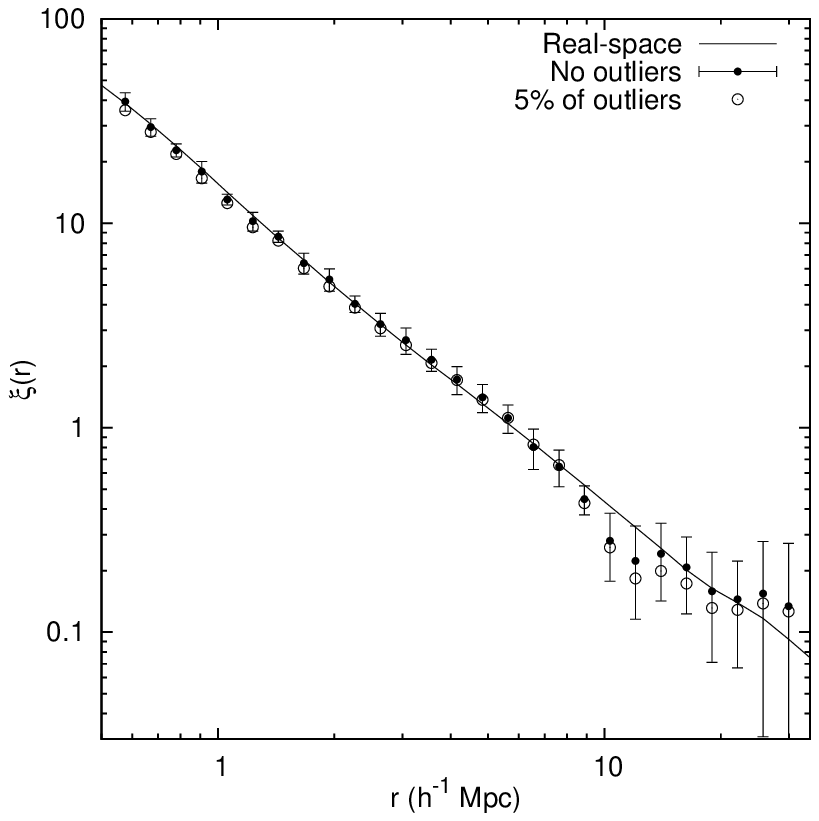}
\caption{The deprojected correlation function obtained for the $\Delta z = 0.015(1+z)$ catalogue with and without outliers.
The continuous line is the real-space correlation function.}
\label{fig:outliers}
\end{figure}

In order to assess the robustness of the method to the presence of `catastrophic' redshift determinations,
we repeated the calculation in catalogues containing 5 per cent of such outliers. Outliers were created selecting points at random
in the original catalogue, and assigning them a random distance within the range considered. Even with the conservative
assumption of a large fraction of outliers, our method recovers $\xi(r)$, 
although the values of $\Delta\xi$ are slightly larger in this case, ranging from 5 to 
13 per cent for $r < 10 \hMpc$ and $\Delta z \leq 0.015 (1 + z)$. The result obtained 
for the catalogue with $\Delta z = 0.015 (1+z)$, and containing 5 
per cent of outliers is shown in Fig.~\ref{fig:outliers}.
Similar results were obtained when a fraction of the outliers were taken from a Poisson sample within our volume.
This case would reproduce the effect of stellar contamination in the catalogue.

%%%%%%%%%%%%%%%%%%%%%
%%%%%%%%%%%%%%%%%%%%%

We performed an additional test of the deprojection method using 
a realization of a segment Cox process, for which an analytical 
expression of the correlation function is known \citep{mar98a}. The results of those tests are 
further explained in Appendix~\ref{sec:app}.

\section{Conclusions}
\label{sec:conc}

We have shown the reliability of recovering the real-space two-point
correlation function from photometric redshift surveys. We have used
light-cone simulations to produce mock catalogues that have been
distorted by randomizing along the line of sight the object positions
following Gaussian distributions with different variances similar to
the associated nominal errors of the photometric redshift surveys
$\Delta z/(1+z)$. 

The method used to recover the real-space correlation function consists in 
obtaining the projected correlation function
by integrating the two-dimensional correlation function along the line of 
sight. The projected correlation function is then deprojected assuming that 
redshift errors do not affect transverse distances.

The deprojection method applied on the distorted mock
surveys provides quite satisfactory results for recovering the
real-space correlation function. We have quantified the quality of
the recovering process as a function of the errors in the
photometric redshifts. Our method was able to recover the real-space
correlation function within a $5$ per cent for $r < 10 \hMpc$ from 
photometric catalogues with $\Delta z \leq 0.015 (1 + z)$. For larger
redshift errors, the method is only valid (within a $7$ per cent) 
for smaller scales, $r < 2 \hMpc$. 
Hence, our method allows the extraction of useful information on the clustering
of galaxies through the correlation function. That information can be used for the estimation
of cosmological parameters based on data from photometric redshift surveys.

We discuss now possible alternatives to the method exposed here. 
A variation of the method would be to use a smaller value of $\pi_{\rm max}$ 
in the integration of equation~(\ref{eq:wp}), and multiply the result by a constant correction factor.
This would reduce the extra noise introduced by the integration along a large
range in the $\pi$ direction.
We found that, for $\pi_{\rm max} \simeq \Delta z$, a correction factor of $\simeq 2$ works well in our
simulation, generally reducing the error. However, the optimal value is slightly different for each mock 
photometric catalogue.  The main problem for the use of this method would be the accurate determination
of the correction factor in each case, as any deviation from the optimal value would introduce a bias in the result. 
The Gaussian approximation used here is probably not so close to reality as to infer that constant from
our simulations.

Another possible alternative to the method described in this work could make use of the
new estimator $\omega$ presented in \citet*{pad07a}. They proposed $\omega$ as
an alternative to the projected correlation function to use with spectroscopic survey
data. 
In principle, it can also be applied to photometric redshift survey data, in a way similar
to the one presented here, but we did not investigate further this possibility.

\section*{Acknowledgments}

We thank our anonymous referee, whose comments helped to improve the 
clarity and quality of the paper. 
We acknowledge
support from the Spanish Ministerio de Educaci\'on y Ciencia (MEC) through
project AYA2006-14056 (including FEDER), from the Estonian Science Foundation
through grant ETF6104, and from the Estonian Ministry for Education and
Science through research projects SF0062465s03 and SF0060067s08. 
PAM acknowledges support from the
Spanish MEC through a FPU grant. IS was supported by CIMO's ETF grant 5347.
PH  was supported by the Jenny and Antti Wihuri foundation. This work has also been  
supported by the University of Valencia through a visiting professorship for Enn Saar.

\bibliographystyle{mn2e}
\bibliography{bibliografia}

\appendix

\section{Test of the method using a segment Cox process}
\label{sec:app}

In order to test our method against an analytical prediction for $\xi(r)$, 
we used a distribution of points 
given by a segment Cox process. This process is produced in the following way 
\citep{mar02a}: segments of a given length,
$l$, are randomly scattered within a volume. Then, points are randomly distributed 
along these segments. The lenght density 
of the system of segments is $L_V = \lambda_s l$, where $\lambda_s$ is the mean 
number of segments per unit volume. 
The density of the point process is then
\[
\lambda = \lambda_l L_V = \lambda_l \lambda_s l \, ,
\]
where $\lambda_l$ is the mean number of points per unit length of the segments. The correlation 
function of the point process equals the correlation
function of the system of segments \citep*{sto95a}, which is given by
\begin{equation}
\label{eq:xi_cox}
\xi_{\rm Cox}(r) = \left\lbrace \begin{array}{ll}
	       \frac{1}{2 \pi r^2 L_V} - \frac{1}{2 \pi r l L_V} \, , & r \leq l \\
               0 \, ,                                                 & r  > l 
               \end{array} \right. \, .
\end{equation}

We simulated a segment Cox process in the same volume considered in the 
rest of this work. The parameters we used
were $l = 50 \hMpc$, $\lambda_s = 2 \cdot 10^{-4} \, h^{3}\, \mathrm{Mpc}^{-3}$ and 
$\lambda_l = 4 \, h\, \mathrm{Mpc}^{-1}$, which result 
in $L_V = 0.01 \, h^{2}\, \mathrm{Mpc}^{-2}$ 
and $\lambda = 0.04 \, h^{3}\, \mathrm{Mpc}^{-3}$. These parameters 
were chosen to  approximately match the density of points and the 
behaviour of $\xi(r)$ in the haloes simulation. 
We considered the catalogue obtained directly from the segment 
Cox process as the `real-space' catalogue. We created three
mock `photometric redshift catalogues' following the same 
procedure and using the same values
for $\Delta z$ as described in Section~\ref{sec:data} . 

%%%%%%%%%%%%%%%%
\begin{figure*}
\begin{center}
\includegraphics[scale = 1]{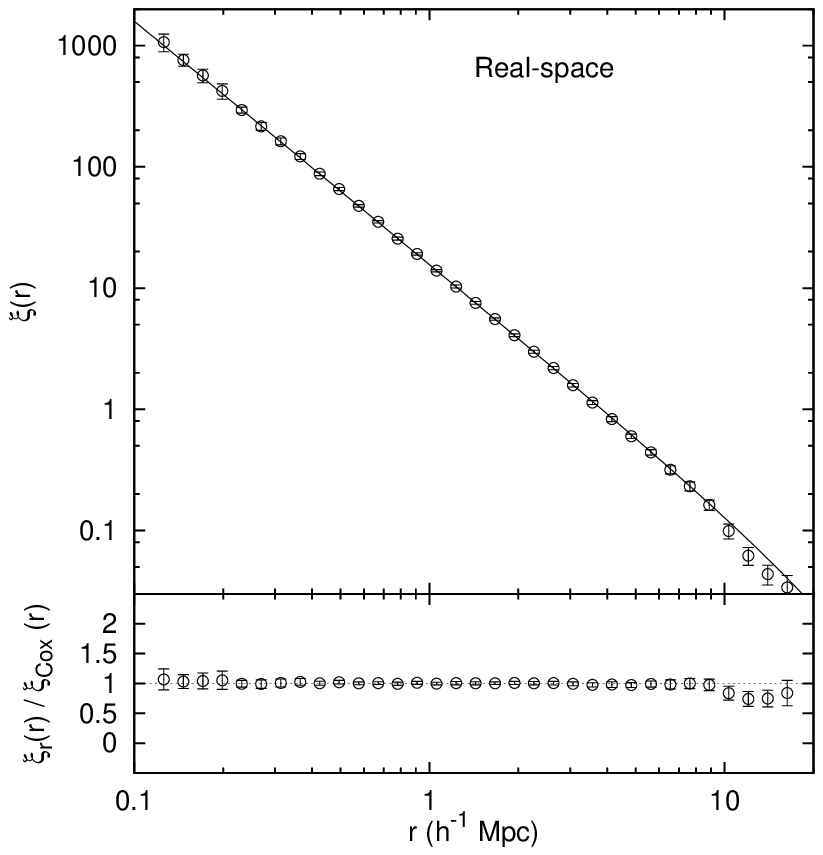}
\includegraphics[scale = 1]{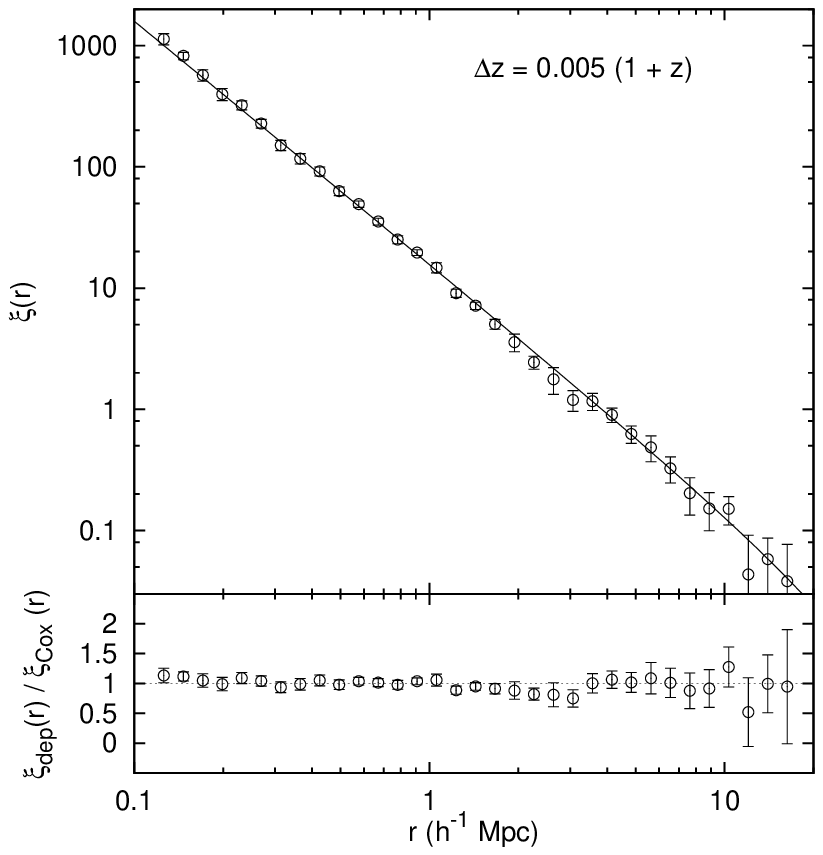}
\includegraphics[scale = 1]{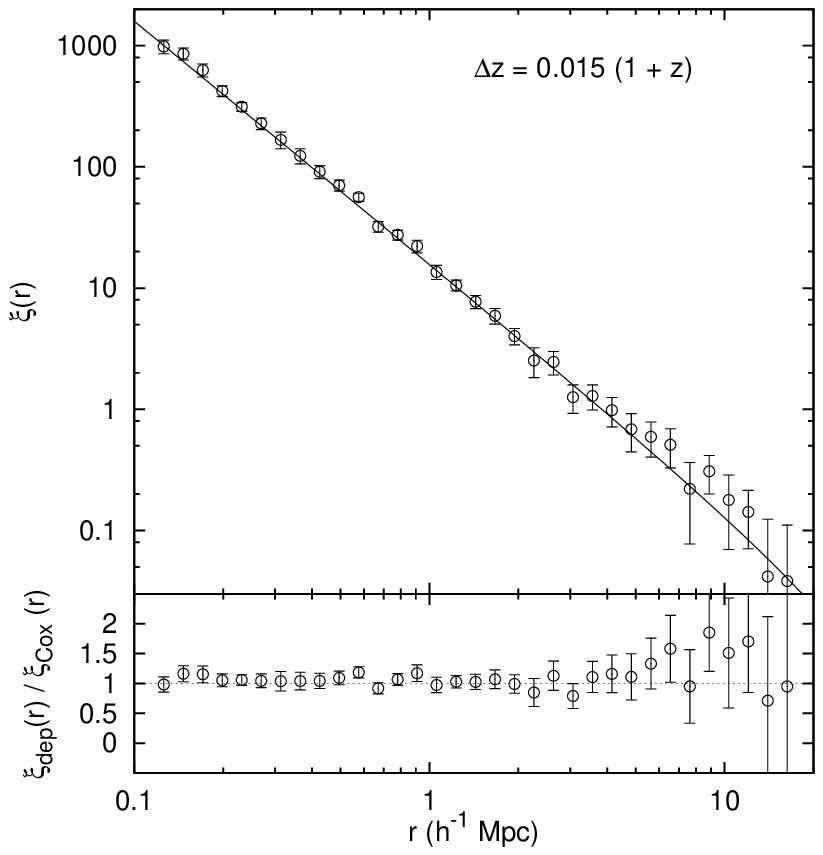}
\includegraphics[scale = 1]{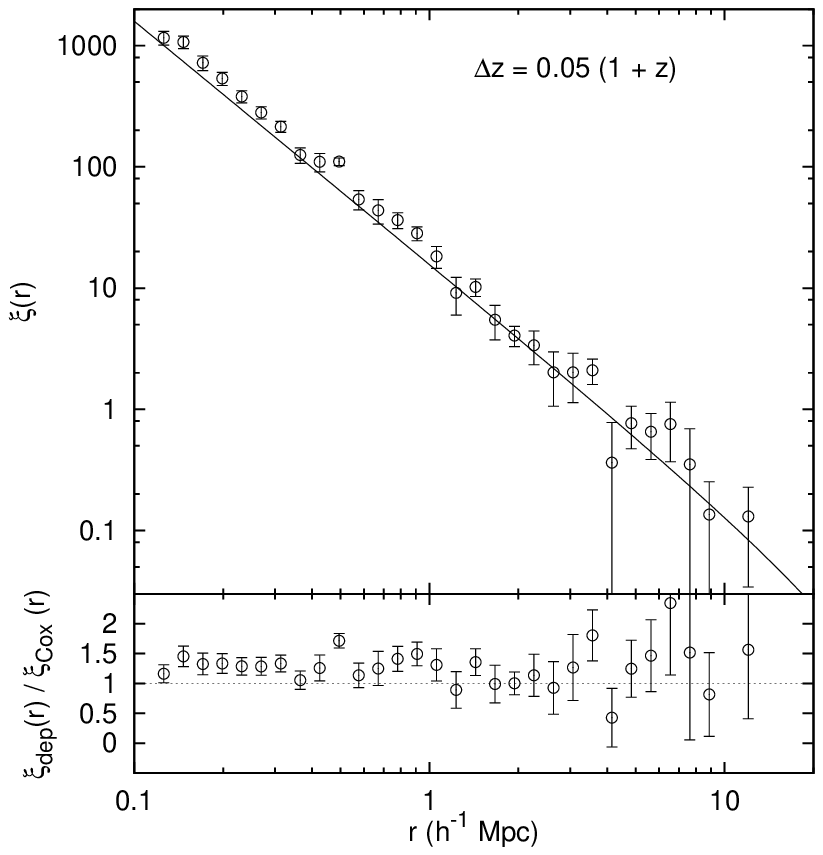}
\end{center}
\caption{The correlation function measured in the real-space and the 
three mock `photometric' catalogues obtained from a segment Cox process 
(open circles), compared to the analytical prediction, equation~(\ref{eq:xi_cox}), 
for this process (solid line).}
\label{fig:xidep_cox}
\end{figure*}
%%%%%%%%%%%%%%%%%

We calculated directly the correlation function for the real-space catalogue according to
equation~(\ref{eq:LS}). 
For the three mock `photometric catalogues',
we used the method described in Section~\ref{sec:meth} to obtain the deprojected 
correlation function. 
The estimation of errors was performed using the same jackknife method 
as described above for the haloes simulation case. We checked that
the errors obtained were comparable to the variance of the results from 
several realizations of the Cox process. 
The comparison of our results to the analytical prediction~(\ref{eq:xi_cox}) 
is shown in Fig.~\ref{fig:xidep_cox}. 

We quantify the quality of the recovery in the same way as we did for the haloes simulation,
using the quantities $\Delta \xi$ and $\widehat{\Delta \xi}$. In this case, 
we define $\Delta \xi$ as the
relative deviation of $\xi_{\rm dep}(r)$ from the analytical prediction 
$\xi_{\rm Cox}(r)$ (equation~(\ref{eq:xi_cox})). 
The values obtained are shown in Table~\ref{tab:DxiCox}.

%%%%%%%%%%%%%%%%%%%%%%%%%5
\begin{table*}
\caption{Values of $\Delta \xi$ and $\widehat{\Delta \xi}$ obtained for the three 
			mock photometric catalogues obtained from a Cox process, and for different scale ranges.}
\label{tab:DxiCox}

\begin{tabular}{@{}lcccccc} \hline
                                      & \multicolumn{2}{c}{$\frac{\Delta z}{(1 + z)}= 0.005 $} & 
										 \multicolumn{2}{c}{$\frac{\Delta z}{(1 + z)}= 0.015 $} & 
                                             \multicolumn{2}{c}{$\frac{\Delta z}{(1 + z)}= 0.05 $} \\ 
Range ($h^{-1}\, \mathrm{Mpc}$) & $\Delta \xi$ & $\widehat{\Delta \xi}$ & $\Delta \xi$ & $\widehat{\Delta \xi}$ & $\Delta \xi$ & $\widehat{\Delta \xi}$ \\ \hline
$0.5 < r  < 10$ &   $0.08$ & $0.16$ & $0.18$  &  $0.23$ &  $0.35$  & $0.40$     \\
$0.5 < r < 2$   &     $0.06$      &  $0.08$      &  $0.07$      & $0.12$  &  $0.23$        &  $0.21$     \\
$2 < r < 10$ &      $0.10$        &  $0.22$    &  $0.27$   & $0.33$   &  $0.46$     &  $0.57$   \\ \hline
\end{tabular}
\end{table*}
%%%%%%%%%%%%%%%%%%%%%%%%%%%%%

We recover the real-space correlation function within a $10$ per cent for the $\Delta z / (1+z) = 0.005$
catalogue. In this case, however, our method starts to fail at $r \simeq 3 - 4 \hMpc$ for the $\Delta z / (1+z) = 0.015$ 
catalogue (this is seen as a larger value of $\Delta \xi$ for this range, and as an increasing trend in Fig.~\ref{fig:xidep_cox}). 
When applying
the method to the $\Delta z / (1+z) = 0.05$ catalogue,  $\xi_{\rm dep}(r)$ is 
consistently higher than $\xi_{\rm Cox}(r)$. Although this bias could be an artefact of this 
particular point process, it also means that the deprojection 
method described in this work can not be fully trusted when it is applied 
to catalogues with large redshift errors.

\bsp

\label{lastpage}

\end{document}